\documentclass{pasj00}
\draft

\newcommand{\casa}{Cassiopeia~A}

\SetRunningHead{The Suzaku \casa team}{Possible keV-TeV correlation in the reverse shock}
\Received{2008/11/21}
\Accepted{2009/07/15}
\begin{document}
\title{Suzaku X-ray imaging and spectroscopy of \casa }
\author{
Yoshitomo \textsc{Maeda},\altaffilmark{1}
Yasunobu \textsc{Uchiyama},\altaffilmark{1,2}
Aya \textsc{Bamba},\altaffilmark{1}
Hiroko \textsc{Kosugi},\altaffilmark{3}\\
Hiroshi \textsc{Tsunemi},\altaffilmark{3}
Eveline A. \textsc{Helder},\altaffilmark{4}
Jacco \textsc{Vink},\altaffilmark{4}\\
Natsuki \textsc{Kodaka},\altaffilmark{5}
Yukikatsu \textsc{Terada},\altaffilmark{5}
Yasushi \textsc{Fukazawa},\altaffilmark{6}
Junko \textsc{Hiraga},\altaffilmark{7}
John P. \textsc{Hughes},\altaffilmark{8}\\
Motohide \textsc{Kokubun},\altaffilmark{1}
Tomomi \textsc{Kouzu},\altaffilmark{5}
Hironori \textsc{Matsumoto},\altaffilmark{9}\\
Emi \textsc{Miyata},\altaffilmark{3}
Ryoko \textsc{Nakamura},\altaffilmark{1}
Shunsaku \textsc{Okada},\altaffilmark{1}
Kentaro \textsc{Someya},\altaffilmark{1}\\
Toru \textsc{Tamagawa},\altaffilmark{7}
Keisuke \textsc{Tamura},\altaffilmark{1}
Kohta \textsc{Totsuka},\altaffilmark{10}
Yohko \textsc{Tsuboi},\altaffilmark{10}\\
Yuichiro \textsc{Ezoe} ,\altaffilmark{11}
Stephen S. \textsc{Holt} ,\altaffilmark{12}
Manabu \textsc{Ishida},\altaffilmark{1}\\
Tsuneyoshi \textsc{Kamae},\altaffilmark{2}
Robert \textsc{Petre},\altaffilmark{13}
Tadayuki \textsc{Takahashi},\altaffilmark{1}
}
\altaffiltext{1}{Department of High Energy Astrophysics,
  Institute of Space and Astronautical Science (ISAS), \\
  Japan Aerospace Exploration
  Agency (JAXA), 3-1-1 Yoshinodai, Sagamihara, 229-8510, Japan}
\altaffiltext{2}{Kavli Institute for Cosmology and Particle Astrophysics, Stanford Linear Accelerator Center, \\
2575 Sand Hill Road M/S 29, Menlo Park, CA 94025, USA}
\altaffiltext{3}{Department of Earth and Space Science,
Graduate School of Science, Osaka University,\\
Toyonaka, Osaka 560-0043, Japan}
\altaffiltext{4}{Astronomical Institute Utrecht, Utrecht University, P.O. Box 80000, NL-3508 TA Utrecht, \\
The Netherlands}
\altaffiltext{5}{Department of Physics, Saitama University, Saitama 338-8570, Japan.}
\altaffiltext{6}{Department of Physical Science, Hiroshima University, \\1-3-1 Kagamiyama, Higashi-Hiroshima, Hiroshima 739-8526, Japan}
\altaffiltext{7}{Cosmic Radiation Laboratory, RIKEN, 2-1 Hirosawa, Wako, Saitama 351-0198, Japan}
\altaffiltext{8}{Department of Physics and Astronomy, Rutgers University,\\ 136 Frelinghuysen Road, Piscataway, NJ 08854-8019, USA}
\altaffiltext{9}{Department of Physics, Graduate School of Science, Kyoto University, \\Kita-shirakawa Oiwake-cho, Sakyo, Kyoto 606-8502, Japan}
\altaffiltext{10}{Department of Physics, Chuo University, 1-13-27 Kasuga, Bunkyo-ku, Tokyo 112-8551, Japan}
\altaffiltext{11}{Department of Physics, Tokyo Metropolitan University, 1-1 Minami-Osawa, Hachioji, Tokyo 192-0397, Japan}
\altaffiltext{12}{F. W. Olin College of Engineering Needham, MA 02492, USA}
\altaffiltext{13}{NASA/Goddard Space Flight Center, Code 662, Greenbelt, MD 20771, USA}

\email{ymaeda@astro.isas.jaxa.jp}
\KeyWords{ISM: individual (Cassiopeia A) - ISM:supernova remnants } 
\maketitle 
\begin{abstract}

Suzaku X-ray observations of a young supernova remnant, \casa, were
carried out. K-shell transition lines from highly ionized ions of
various elements were detected, including Chromium (Cr-K$\alpha$ at
5.61 keV).  The X-ray continuum spectra were modeled in the 3.4--40
keV band, summed over the entire remnant, and were fitted with a
simplest combination of the thermal bremsstrahlung and the non-thermal
cut-off power-law models. 
The spectral fits with this assumption
indicate that the continuum emission is likely to be dominated by the
non-thermal emission with a cut-off energy at $>$ 1 keV. 
The
thermal-to-nonthermal fraction of the continuum flux in the 4--10 keV
band is best estimated as $\sim$0.1. Non-thermal-dominated
continuum images in the 4--14 keV band were made. 
The peak of the
non-thermal X-rays appears at the western part.
The peak position of the TeV $\gamma$-rays measured with
HEGRA and MAGIC is also shifted at the western part with the 1-sigma
confidence. 
Since the location of the X-ray continuum
emission was known to be presumably identified with the reverse shock
region, the possible keV-TeV correlations give a hint that the
accelerated multi-TeV hadrons in \casa\ are dominated by heavy
elements in the reverse shock region.

\end{abstract}

\section{Introduction}

Supernova remnants (SNRs) have long been considered to be the primary
acceleration sites of cosmic-ray particles below the energy of the
so-called {\it knee} in the cosmic ray spectrum, $\sim 10^{15}~{\rm
  eV}$.  The first evidence for multi-TeV acceleration was the
discovery of synchrotron X-ray emission from the shell of the
supernova remnant SN~1006 \citep{Koyama95}. More direct evidence for
the presence of multi-TeV particles (electrons and/or protons) inside
supernova remnants was provided by CANGAROO's discovery of TeV
$\gamma$-rays from another and brighter shell-like SNR in TeV, RX
J1713.7$-$3946 \citep{muraishi00}.  
Recent high quality morphological
and spectral studies with the HESS TeV imager combined with the X-ray
imager such as ASCA, found a good keV-TeV correlation from RX
J1713.7$-$3946 (\cite{Aharonian04}) and RXJ 0852.0$-$4622
(\cite{Aharonian07}), indicating that both the X-rays and the TeV
$\gamma$-rays are emitted by the TeV particles in the SNR shell. 
The
extremely thin X-ray filaments with a 0.02pc width in SN~1006
\citep{Bamba03,long03}, as well as the X-ray variability on time
scales of a year from RX J1713.7$-$3946 (\cite{Uchiyama07}) strongly
support the efficient acceleration of particles by SNR shocks. Up to
now, many observational results support that supernova remnants (SNRs)
are major sources of galactic TeV cosmic rays.

The young ($\sim 330\ \rm yr$ old) supernova remnant \casa\ is one of
several SNRs from which non-thermal X-rays and TeV
$\gamma$-rays have both been detected (X-rays:\cite{Allen97,Uchiyama08a},
TeV:\cite{Aharonian01,Albert07}). 
 In X-rays
\casa\ seems to consist of a number of 
thermal and non-thermal X-ray emitting knots/filaments
(\cite{Hughes00,Hwang04,Bamba05}). Although some non-thermal emission
is associated with the forward shock, 
the dominant
source of non-thermal emission may be identified with the
reverse shock regions (\cite{Helder08} and references there in).
It therefore is a unique object in which we can
study the particle acceleration by the reverse shock, because for the
other SNRs the acceleration seems to originate from 
the forward shock region only (e.g., \cite{Parizot06}).
The bulk of the composition of the high-energy cosmic rays ($\sim$ PeV) might be made up of elements heavier than  helium (\cite{Tibet06} and reference there in). The study of the acceleration in the reverse shock is interesting since the ejecta is dominated by these heavy elements.

The overall X-ray spectra from \casa\ are described by optically thin-thermal
emission (line emission by highly ionized heavy elements and
bremsstrahlung) accompanied by non-thermal emission extending beyond 10
keV (e.g., \cite{Serlemitsos73,Pravdo79,Allen97}).
However,
the basic characterization of the two continuum components---such as the total
intensity, spectral shapes (temperature of thermal bremsstrahlung and
the shape of synchrotron spectrum), and the emission sites---has been
somewhat controversial.  For example, \citet{Allen97} fitted the
broadband spectrum obtained with RXTE using thermal bremsstrahlung
with $kT_e = 2.9$ keV and a broken power law with $\Gamma_1 = 1.8$ and
$\Gamma_2 = 3.0$ with a break energy of 15.9 keV, while \citet{Favata97} fitted the Beppo-SAX spectrum of \casa\ with $kT_e = 3.8$
keV and a power law with $\Gamma = 2.9$ (see also \citet{Vink03}).

In this paper, we present a Suzaku study of the X-ray
emission from \casa.  
The X-ray Imaging Spectrometer (XIS) enables us to make hard
X-ray images and spectra up to $\sim 14~{\rm keV}$.
By combining the XIS with
the companion instrument, Hard X-ray Detector (HXD), we present a
wide-band continuum spectrum (3.4--40~keV).

\section{Observation and Data Reduction}

The Suzaku satellite \citep{mitsuda07} has carried out three
observations of \casa, one in 2005 September, and two in 2006
February.  The observation log is given in table~\ref{tabl:obslog}.  
In all
the three observations, \casa\ was pointed at around the geometrical center of the XIS detector (the XIS nominal position). 

\begin{table}[hbt]
  \begin{center}
\caption{Suzaku Observations of \casa}
\label{tabl:obslog}
    \begin{tabular}{lcccc}
\hline \hline
 No. & Date             & \multicolumn{2}{c}{Exposure (ks)} & XIS mode \\ 
          & YYYY/MM/DD       & XIS       & HXD              & Clock/Window/Burst/S-CI$^{b}$  \\ \hline
 1st       & 2005/09/01        & 30.3$^{a}$  & 20.5           & Normal/-/-/N  \\
 2nd       & 2006/02/02        & 0           & 8.6            & Normal/-/4sec/N  \\
 3rd       & 2006/02/17        & 7.1        & 17.1           & Normal/-/4sec/N \\ \hline
   \end{tabular}
  \end{center}
$a$: Only the data taken in segment B are available. Those in other segments (A,C,D) are mostly suffered with telemetry saturation. \\
$b$: Clock/Window/Burst/S-CI are the options of the CCD clocking or editing modes. See \citet{koyama07a} for details. 
\end{table}

\medskip

Suzaku's XIS (\cite{koyama07a}),
consists of four X-ray CCD cameras each placed in the focal plane of
the X-Ray Telescope (XRT: \cite{serlemitsos07}).  All four XRT modules
are co-aligned to image the same region of the sky, with a field of view of
18\arcmin$\times$18\arcmin. The point spread function has
a half power diameter of 1\farcm9--2\farcm3.  
Three of the cameras (XIS\,0, XIS\,2, and XIS\,3)
have front-illuminated (FI) CCDs, sensitive in the 0.4--14~keV energy
band, and the remaining one (XIS\,1) has a back-illuminated (BI) CCD,
which is
sensitive down to 0.2~keV.  
For a broad band fitting of the continuum emission, we analyzed the FI
CCDs (XIS\,0\,2\,3), since the hard X-ray data above $\sim$8~keV is
noisy in the BI CCD XIS\,1. 
Combined with XRT, the total effective
area per one XIS system is $\sim160$~cm$^2$ at 8~keV.
Suzaku also has a non--imaging Hard X-ray Detector (HXD:
\cite{kokubun07, takahashi07}).  The HXD is comprised of Si PIN diodes
(PIN) sensitive in the 15--80~keV range, and a 
GSO scintillator (GSO) sensitive in the 
50--500~keV range. Both are located inside an active BGO shield.  The PIN and
GSO have a field of view (FOV) of 34\arcmin$\times$34\arcmin\ and
4\degree$\times$4\degree , respectively, with the lowest non--X-ray
background ever achieved.

The XIS and the HXD data were processed with the Suzaku pipe-line
software (version 2.1).  The response functions were generated by
using the CALDB 2008-07-09.  We removed data taken during the South
Atlantic Anomaly passages and data for which earth elevation angles were below
5\degree.  For PIN and GSO data, we further excluded the events taken
with cut-off rigidity less than 6~GV.  We corrected for the dead time
$(\sim4$--8\%) of the HXD using \texttt{hxddtcor}.  
The total net
exposure time after this filtering is summarized in table
\ref{tabl:obslog}. The exposure for the HXD (PIN and GSO) is
$46$~ks.  The exposure time for XIS is given below.

In figure~\ref{tiled}, we show an example of the \casa\ XIS-0 image taken at the 1st and 3rd
observations. No XIS data are available at the 2nd observations because
of an accident in the XIS operation. The image of the 1st observation
is truncated at the south-east rim while that of the 3rd is not. The
truncation is due to the data lost by the telemetry saturation.  The
location of the image truncation is different from detector to
detector in order to its orientation (see \cite{ishisaki07}).

An amount of the events recorded in the XIS readout interval (8
second) is limited (\cite{koyama07a}). The event rate at the 1st
observation was slightly above the telemetry limit. Consequently, a
small fraction of the data were lost; $\sim$15\% for the FI CCDs but
varied from interval to interval. The larger fraction of the data
($\sim30$\%) was lost in the BI CCD (XIS-1) than in the FI CCDs.  The
read-out of the CCD chip is separately made at each quarter called as
`segment'. The data recording to the telemetry in the read-out
interval is then made with a scheduled order of
B$\rightarrow$C$\rightarrow$A$\rightarrow$D. The telemetry saturation
for our data is usually occurred during the read-out of the data in the
segment C in which roughly the half of the \casa\ events are
detected. All the data in the segment B containing another half of the
\casa\ image are fully recorded but is hardly in the D. It is in
general impossible to recover the data lost by the telemetry
saturation. We gave up to use the telemetry saturated data for the
image analysis (Section 3.1).  We also did not use them for the
XIS-HXD combined analysis (Sections 3.2 and 3.3) for which their
relative normalization is crucial.  The XIS exposure for the
non-telemetry-saturated data of the 3rd observations is only
$\sim$7~ks long after being corrected for the dead-time effect
discussed at the end of this section.

It is notable that a large fraction of the events from \casa\ were
recorded in the 1st observation. We therefore used the
telemetry-saturated data for the Cr-K line analysis because an energy ratio
or a flux ratio between the two lines are less dependent on the weak
telemetry saturation. Since the exposure time of the 1st observations
is 30 ksec long, their statistics are four times better
than those of the 3rd. The spectra made with the high statistics of
the 1st observation were used in section 3.4 and indeed help
to restrict the line parameters precisely.

At the 3rd observation, the XIS detector was made with the 4-sec burst
option. The burst option selects the exposure time arbitrarily (in
1/32s steps) within the readout interval of 8 s and provides a
full-sized image. This option introduces an artificial dead time.  For
the 4-sec burst option, the exposure is 4 sec and a dead time of 4 sec
is introduced every 8 sec. The amount of the events in the interval is
reduced by a factor of two. 
The burst option also works to reduce
the numbers of the pile-up events. The pile-up issue will be discussed
in the section below.

\begin{figure}[!htb]
  \begin{center}
1st\hspace{5cm}3rd\\
\includegraphics[width=16cm,angle=0]{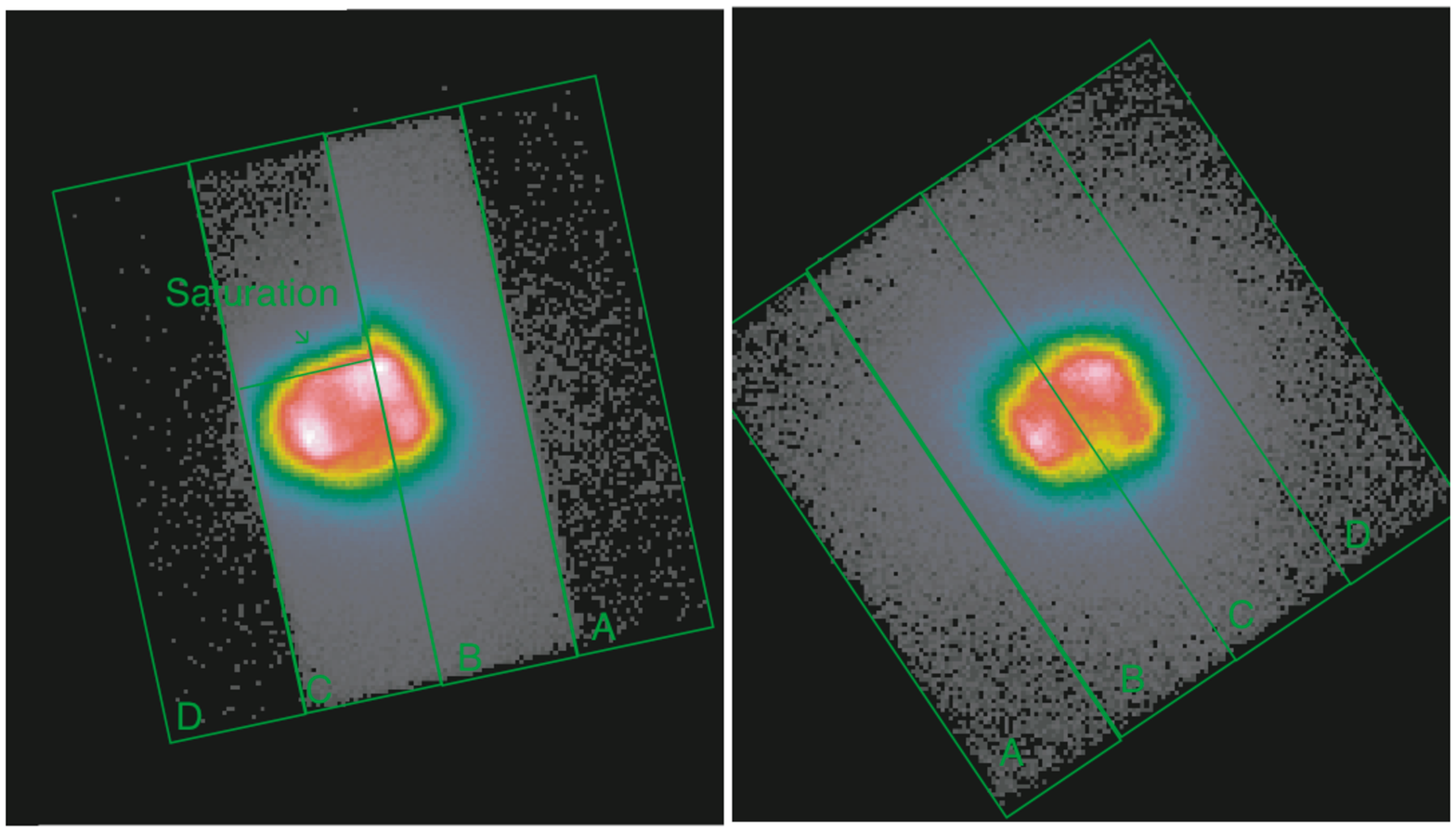}
  \end{center}
  \caption{XIS-0 images of the 1st and 3rd observations. Each segment is outlined with square.  }
  \label{tiled}
\end{figure}


\begin{figure}[!htb]
  \begin{center}
    \FigureFile(135mm,135mm){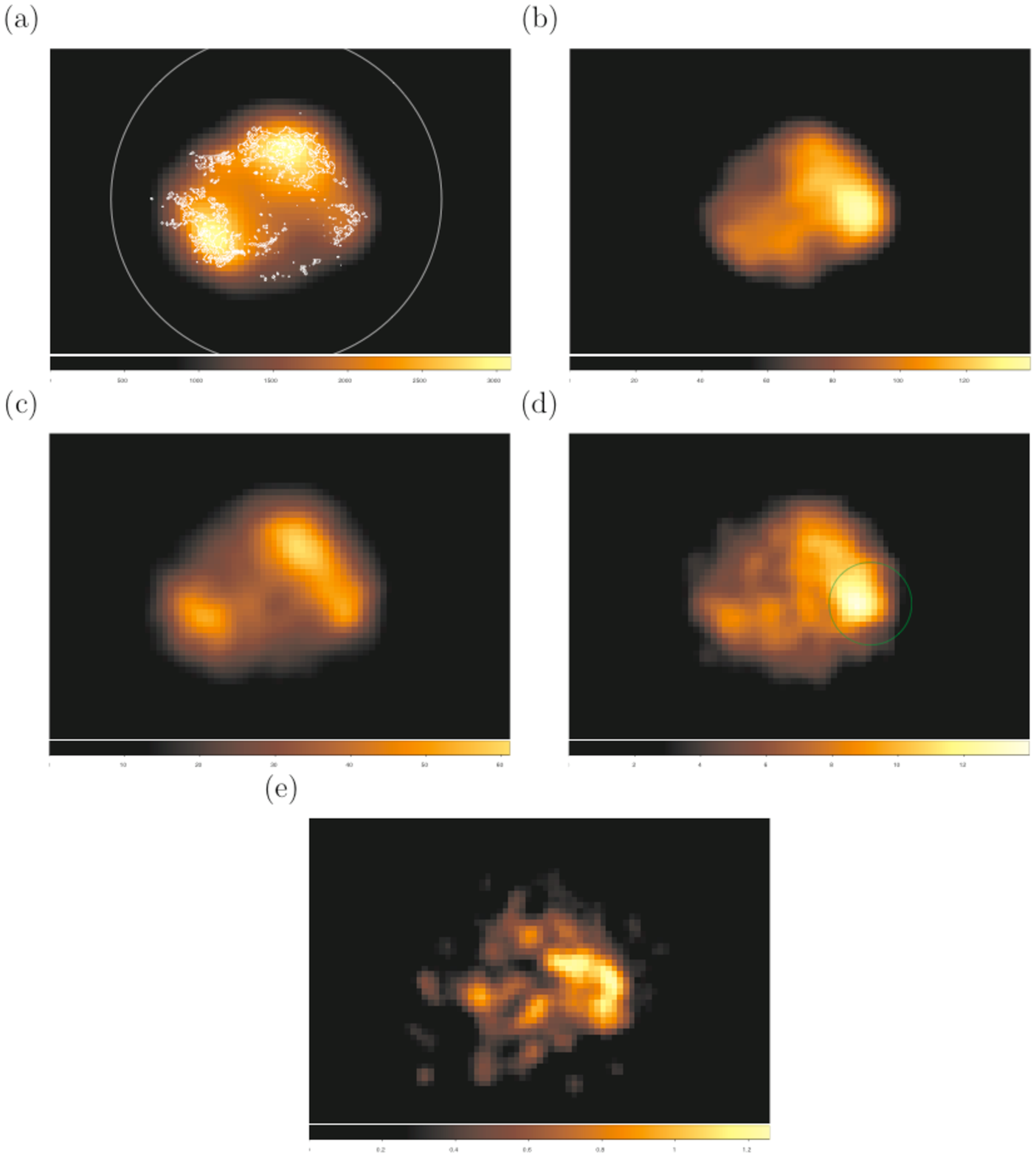}
  \end{center}
  \caption{XIS images taken at the third observation in the five
    band: all (0.5--14 keV), 4--6, 6--7, 8--11, 11-14 keV.  The white contour in
    figure (a) is made using the Chandra ACIS image.  The green circle
    in figure (d) is the region that we take the spectra for the
    western spot (table~\ref{tabl:localspot}). The green circle is 2 arcmin in diameter centered at $($R.A.,Dec.$)=(350.7961,58.8099)$.  The while circle in figure
    (a) corresponds to the region we take the overall spectra of
    \casa(table~\ref{bestfit}). The color bars are in units of counts/bin. }
  \label{xisimages}
\end{figure}

\section{Analysis and Results}

\subsection{Images}

Figure~\ref{xisimages} shows the energy-resolved images with the XIS
in 4.2--6, 6--7, 8--11 and 11--14 keV band, respectively. The images
were made adding the XIS\,0, \,2 and \,3 data taken at the third
observations in 2006 February 17th.  
The images are not corrected for
the exposure map, i.e., the vignetting function of the telescope.  The
exposure map in the sky region where \casa\ was focused, is constant
within $\sim$20\% at any energy.

We first made smoothed images of the Suzaku XIS and Chandra ACIS
in the 0.5--14 keV band. The Suzaku images were binned by $8\times8$
pixels and were then smoothed with a Gaussian function with a sigma of 3
bins
($=0.42$~arcmin)
.  The ACIS image was downloaded from the supernova archive web
site \footnote{http://asc.harvard.edu}.  We then aligned the peak
position at the south-east rim of the XIS image to that of the ACIS,
since the Suzaku position has the systematic error of $\sim$20~arcsec
(\cite{Uchiyama08b}). The image after the position correction was
shown in figure~\ref{xisimages}. The XIS image is well matched with
the ACIS contour.

\begin{figure}[!htb]
  \begin{center}
\includegraphics[width=8cm,angle=270]{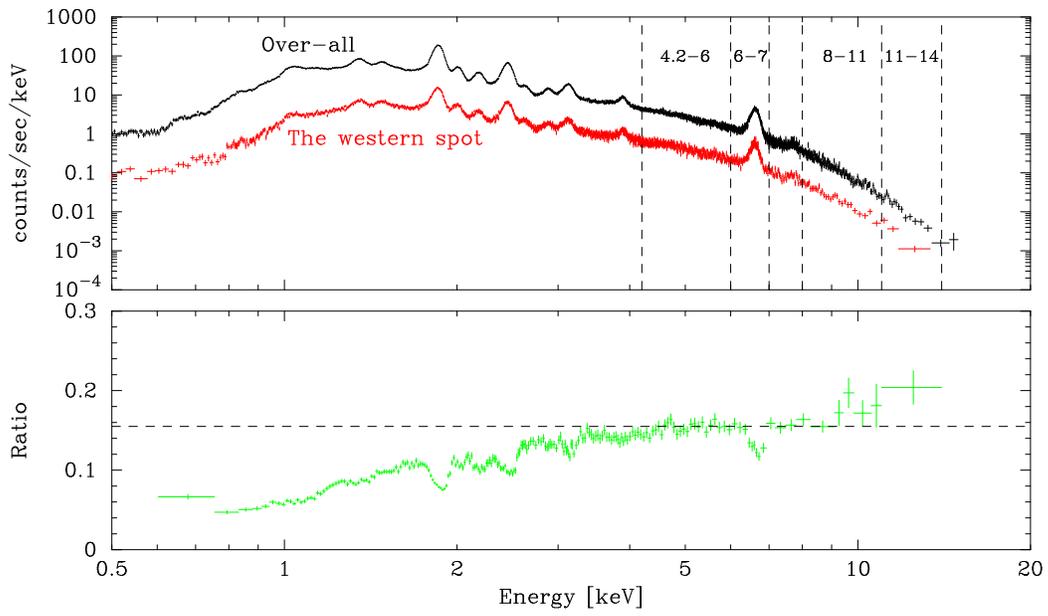}
  \end{center}
  \caption{XIS spectra at the third observations. The black and red data correspond to the overall and the western-spot spectrum, respectively. The lower panel shows their ratio. }
  \label{xisspec}
\end{figure}

By referring to the XIS spectrum (figure~\ref{xisspec}), we confirmed
that the 6--7 keV band is dominated by line emission from 
ionized iron, while the 4.2--6, 8--11 and 11-14 keV bands are
dominated by continuum emission.  The thermal-dominated 6--7 keV band image
shows the brightest peak at the northern part, while others do at
the western part. 
This view looks consistent with that given by
\citet{Bleeker01} using the XMM-Newton data. 
Above 4 keV the events are biased toward the soft energy
border of a given energy band, due to the decline in the effective area and
due to the fall-off in the source spectrum  toward higher energies
(figure~\ref{xisspec}). Therefore,
the images in the 4.2--6, 8--11 and 11--14 keV bands can be eventually
interpreted as images at 4.2, 8, and 11 keV,
respectively.

There may be some concern that  pile-up affects the images, since
\casa\ is known as a very bright X-ray source (118 c/s/FI).
Pile-up results in the identification of two or more photons, as
one photon with a higher energy. It therefore results
in a hardening of a spectrum. 
The pileup is best visible in the 8-11 and 11-14 keV bands because of
their intrinsically low count rates.
We checked the hardening effect of the pile-up
by comparing the spectra of the two observations taken with 
different modes: the normal and the 4-sec burst option.  In the first
observations, taken half a year before the third observation, the
brightest spot at the inner western shell (hereafter the western spot)
is located in segment B of the XIS\,0 and \,3 detectors. Since
the data in the segment B did not suffer telemetry
saturation, we can make a spectrum on the western spot with
them. Table~\ref{tabl:localspot} summarized the count rate and the hardness
ratio obtained with two observations.

The observation with the normal option is twice as sensitive to 
pile-up than those with the 4-sec burst option, because the exposure
time per readout is two times longer. 
The pile-up events make a
continuum-like spectrum. Therefore, the count rate of the energy band
dominated by line emission , i.e., iron 6--7 keV band, is less
increased by the pile-up, but that of the continuum is.  
A larger value of the hardness ratio to the Fe-line band, due to 
hardening, is expected for pile-up affected data. 
However, the hardness
ratio shown in Table~\ref{tabl:localspot} is consistent with being constant for the two modes.
This strongly
indicates that the XIS data are not heavily piled-up and that the flux
in the western spot is stable. 
We thus conclude that the
XIS images are not affected by pile-up at any energy.
 
About 16\% of counts in the 8--11 keV band are located in the region
of the western spot. A similar fraction is obtained for the 4.2--6
and 11--14 keV bands (see also figure~\ref{xisspec}). 
This is roughly consistent with the Chandra ACIS data, for
which we 
find that 19\% of the counts in the 4.2--6 keV are
located in this region.
Therefore, one fifth of the 8--11 keV flux from \casa\ likely
originates from the western spot.

\begin{figure}[!htb]
  \begin{center}
\vspace*{-2cm}
\hspace*{-7cm}
    \FigureFile(125mm,125mm){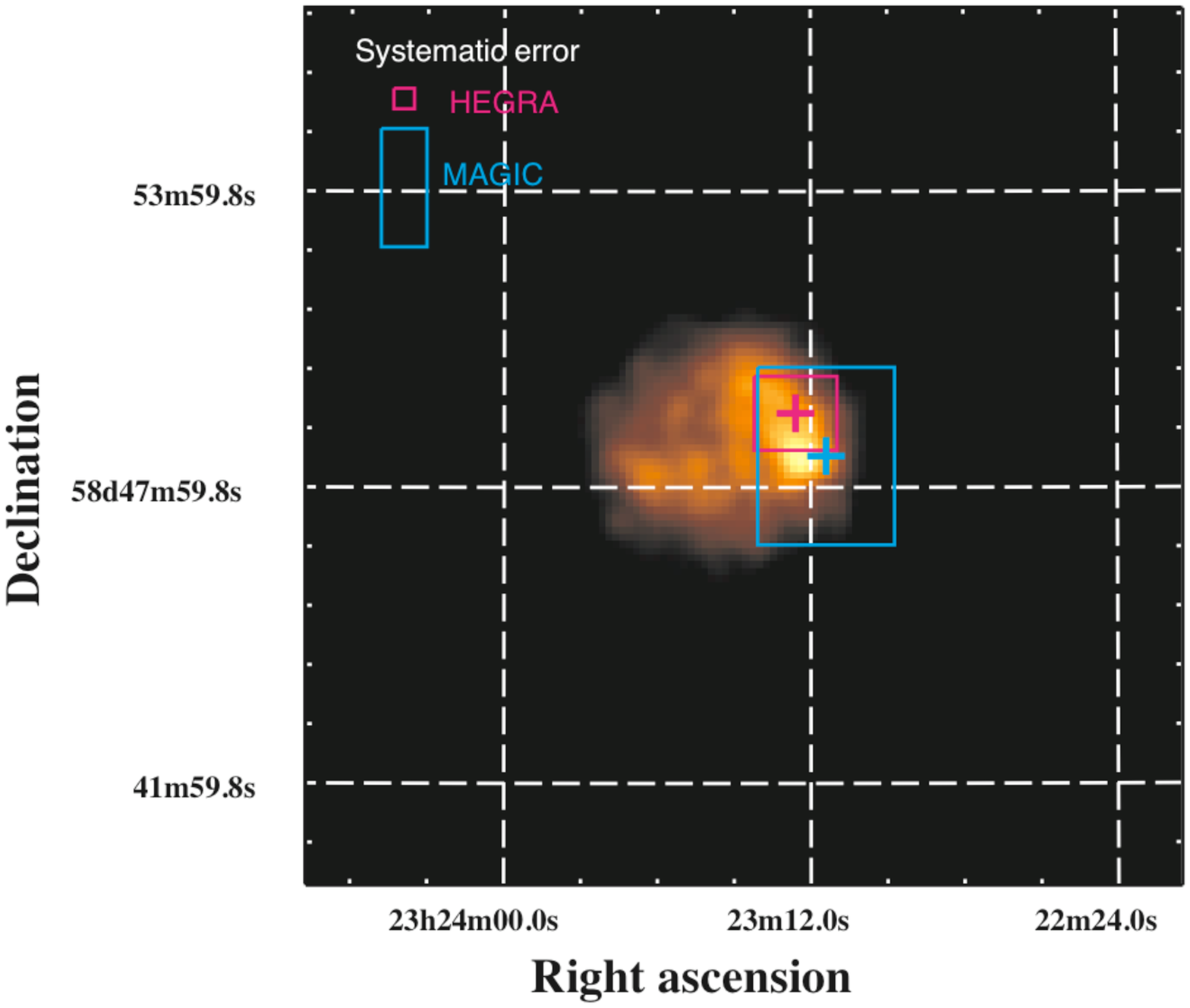} 
  \end{center}
  \caption{Smoothed XIS image in the 8--11 keV band overlaid with the TeV peak position. 
The cross and its surrounding box correspond to the peak position of
the TeV source and its statistical error, respectively: HEGRA (pink) and MAGIC (cyan). The systematic uncertainties of the TeV positions are displayed at the top-left corner. 
The right ascension and declination are written in J2000.  }
  \label{xistev}
\end{figure}

%
Figure~\ref{xistev} shows the 8--11 keV band image overlaid with the
peak positions of the TeV $\gamma$-rays reported by HEGRA
(\cite{Aharonian01}) and MAGIC (\cite{Albert07}).  The confidence
level of the error boxes is 1 sigma. The systematic position
uncertainties (errors) are not included in the boxes, but are displayed
at the top-left of the figure~\ref{xistev}.

Since the other continuum bands of 4.2--6 and 11--14 keV show the
image similar to the 8--11 keV band (Figure~\ref{xisimages}), the
Suzaku image peak of the continuum emission is all shifted to the western
part. The TeV peak measured by the two independent TeV telescopes (HEGRA
and MAGIC) is also shifted to the western direction within its 1-sigma
error. The facts give a hint of a possible keV-TeV correlation in \casa.

\begin{table}
  \caption{Count rate and the hardness ratio of the western spot.} 
\label{tabl:localspot}
  \begin{center}
  \begin{tabular}{ccccc} \hline \hline
\multicolumn{2}{l}{Energy band [keV]}   &       6-7  &     8-11/6-7  &      11-14/6--7 \\ \hline
Detector & Observations & [c/s] &  &  \\
XIS\,0 & & & & \\ 
 & 1st (normal)  & 0.280(3) & 0.24(1) & 0.020(2) \\
 & 3rd (4sec burst) & 0.286(6) & 0.23(1) & 0.018(3) \\
XIS\,3 & & & &  \\
 & 1st (normal)  & 0.263(3) & 0.24(1) & 0.022(2) \\
 & 3rd (4sec burst) & 0.291(6) & 0.24(1) & 0.026(4) \\ \hline
  \end{tabular}
  \end{center}
One sigma Poisson errors are given in parentheses.
\end{table}

\subsection{Spectra}

\begin{figure}[!htb]
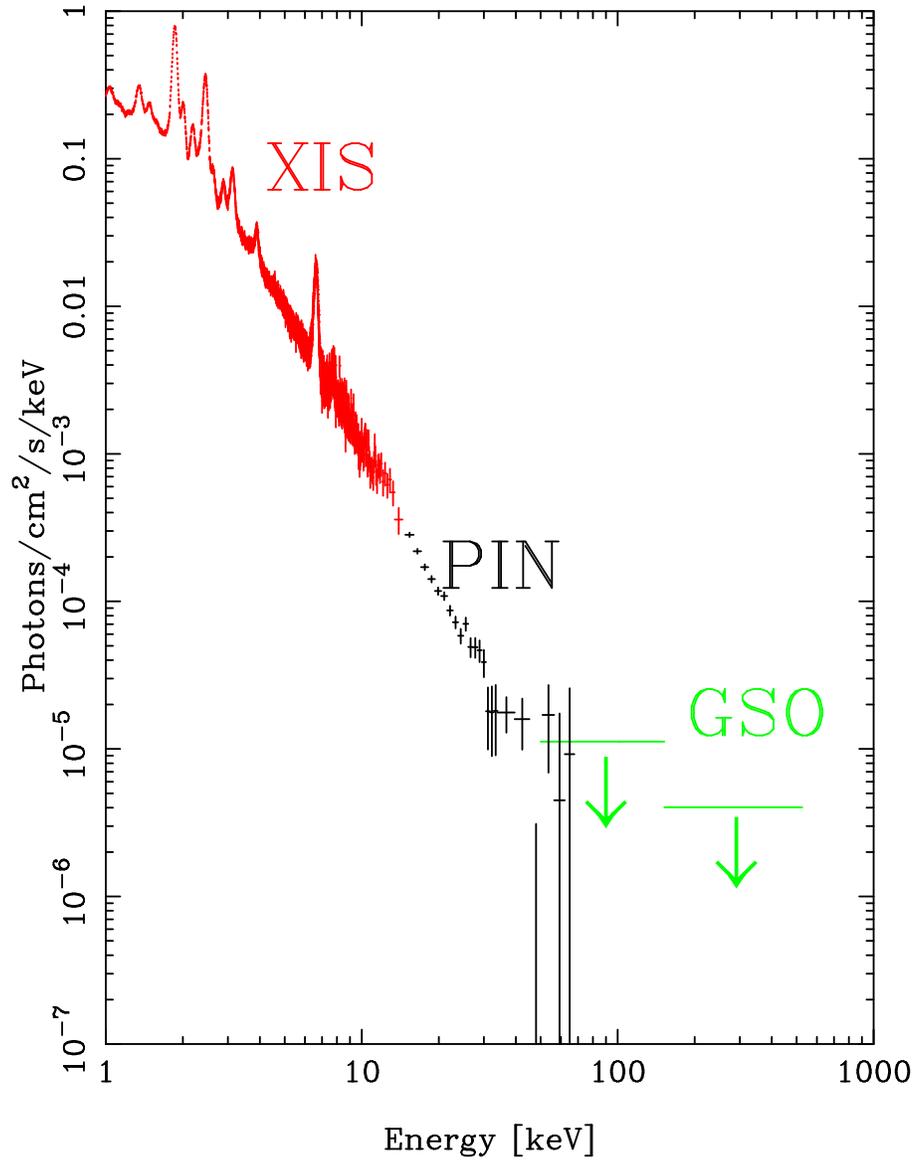

  \begin{center}
    \FigureFile(120mm,80mm){figure4.ps}
  \end{center}
  \caption{Unfolded Suzaku spectrum of \casa.  The error bars for the
    XIS and PIN data are 1$\sigma$, while those for the GSO are 5$\sigma$ upper limit. 
}
  \label{sed}
\end{figure}

Figure~\ref{sed} shows the unfolded Suzaku XIS$+$HXD spectrum in the
1--500 keV band.  The XIS spectrum was made by averaging the XIS\,0,
\,2 and \,3 data of the third observation.  It was also extracted
from the circular region with a radius of 4~arcmin that corresponds to
the circle shown in figure~\ref{xisimages}. The background spectra
were extracted from an annular region with inner and outer radii of
4~arcmin and 7~arcmin. Since the X-ray emission from \casa\ is very strong,
the background contribution is almost negligible.  The background fraction is
only 6 \% above 10 keV, where the background count-rate is the highest.
Therefore the XIS spectra are not sensitive to background
subtraction at any energy range.

Since the HXD PIN and GSO have no imaging capability, we made the
PIN and GSO spectrum by simply averaging the spectra of the three
observations.
We then applied the model background spectra
"tuned non-X-ray background" 
(ver 2.1)
that are released for
every observation
\footnote{http://www.astro.isas.jaxa.jp/suzaku/analysis/hxd/}.
The model background is dominated by
the non X-ray background and does not take into account the cosmic X-ray
background. However, the cosmic X-ray background is negligible for our
spectral analysis
 ($\sim0.15$\% of the source flux).
The count rate
after the background subtraction is reduced down to 38\% and 1.2\% for
PIN (15--80 keV) and GSO (50--500 keV), respectively.

It is well known that the error of the HXD background model
is dominated by systematics. For checking the systematic
uncertainties, we usually compare the background model with data
taken during the earth occultation. However, \casa\ is located so far to the
north that the occultation seldom happened during our observations. It is
very difficult to estimate the systematic error for our particular
background models.

\begin{figure}[!htb]
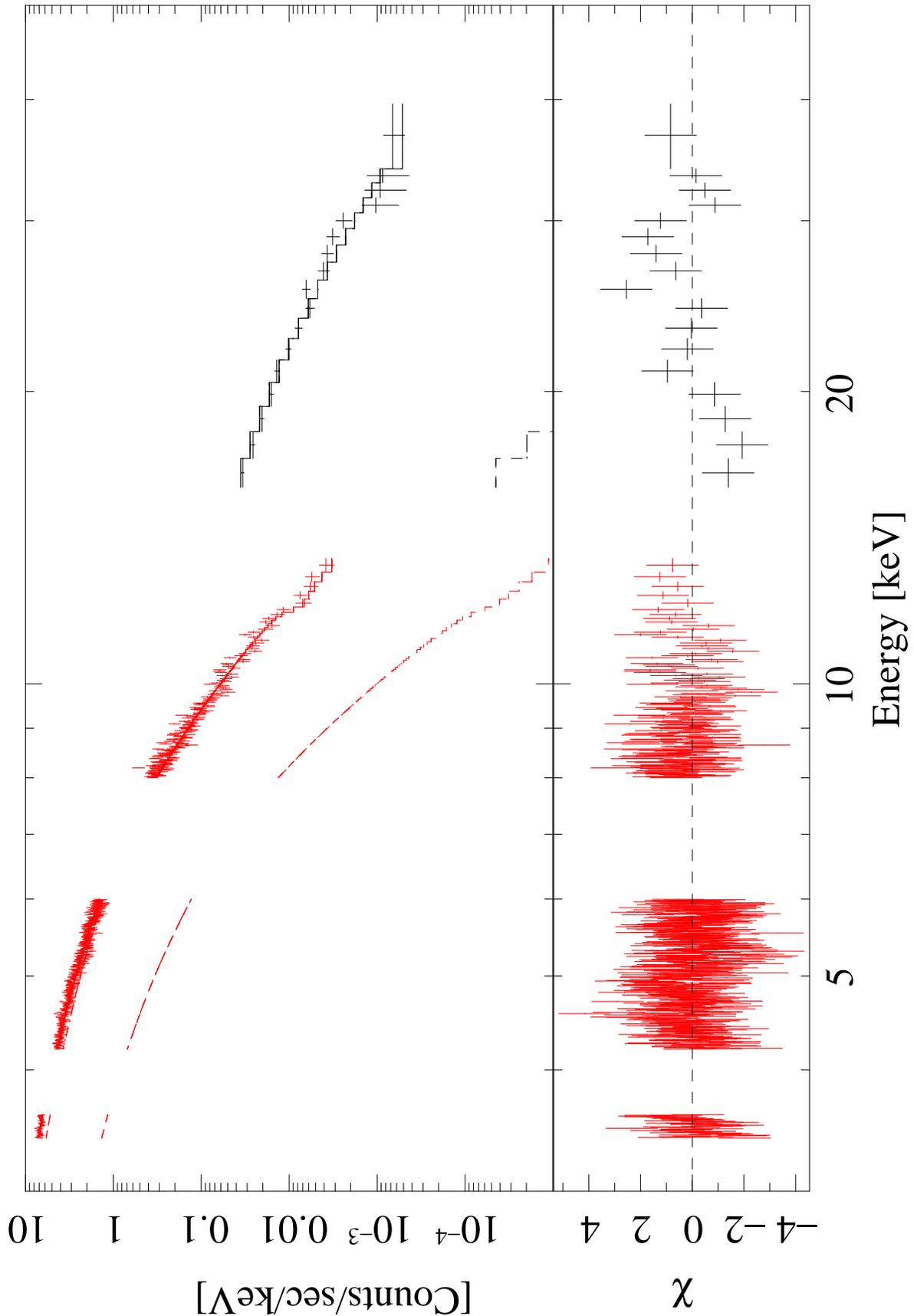

  \begin{center}
    \FigureFile(160mm,160mm){figure5.ps}    
  \end{center}
  \caption{Best-fit models for the Suzaku XIS/PIN spectrum. The solid line
    corresponds to the best-fit parameters of the cut-off power-law model with thermal bremsstrahlung.  
The thermal and non-thermal components of the model are also drawn with dahsed lines. The non-thermal component dominates the flux. 
The bottom panel shows the residuals in
terms of sigmas with error bars of size one.
}
  \label{averagespec}
\end{figure}

We, therefore, refer to the reports on the systematic error based on other
observations provided at the Suzaku web page.  From \citet{Fukazawa09}, we adopted a systematic error of
3\% for the PIN background spectra and 0.6\% for
the GSO (1$\sigma$), and applied them to our analysis. 
The systematic error is included in the error bar of each PIN spectral bin in Figures~\ref{sed} and \ref{averagespec}.
The PIN spectrum is less sensitive to the
systematic error of the background model.

After background subtraction, we found no GSO bins in the spectra
with detections that are statistically significant. 
We thus display 5-sigma upper limits in figure~\ref{sed}. 
The 5-$\sigma$ upper limit is
$1\times10^{-9}$ erg/s/cm$^2$ in the 150--500 keV band by assuming a
photon index of unity. We did not use the GSO data for the spectral
analysis discussed below.

The
over-all spectra from \casa\ consists of thermal and non-thermal components.
In fact, the emission
lines from the thermal component appear in the XIS band. The emission
lines are known to show a variety of  Doppler shifts and are
difficult to be precisely modeled 
(\cite{Holt94,Willingale02})
. Since our paper
mostly focuses on the non-thermal component, we ignored the energy
band dominated by the line emission and 
limited our analysis to the 
continuum-dominated energy bands of 3.4--3.6 and 4.2--6 and 8--14 keV
from the XIS spectra.

The S/N ratio of the HXD PIN spectra gradually drops at around 40
keV.  
The counts in all the PIN bins in the 15--40 keV band are three times or more  as large as the systematic error of the background.
Consequently, for the HXD PIN we limited our
analysis to the 15-40 keV band, which we analyzed jointly with
the continuum dominated bands 
3.4--3.6, 4.2--6, and 8--14 keV of the XIS detectors.

Figure~\ref{averagespec} shows the best-fit parameters of the model
fitting. 
According to a recent calibration report, the PIN normalization of
the Crab nebula is $\sim$15\% higher than that for the XIS.
\footnote{http://www.astro.isas.jaxa.jp/suzaku/doc/suzakumemo/suzakumemo-2008-06.pdf}.
In our spectral fit we, therefore,  multiply the normalization
constant for the PIN by $115\%$. 

The PIN net count rate in the 15--40 keV band is 0.18 c/s that is 63\% of
the background. Since the systematic error of the PIN background is
about 3\%, the normalization uncertainty due to the background
subtraction becomes as large as about 5\%.  The normalization error
can give further uncertainty to the best-fit parameters of our model-fitting.
We evaluated the uncertainty later in this section. 

The continuum consists of both thermal and non-thermal components.
We modeled the thermal component with a thermal bremsstrahlung
emission.  For the non-thermal component, we tested three models, 
1) Power-law, 2) SRCut, and 3) Cut-off power-law. 
For the SRCut model, we fixed the
spectral index at 1 GHz as 0.77 (see \cite{Green04}). 
All the three models show a similar reduced-$\chi^2$ value of 1.15 
(table~\ref{bestfit}). 

The best-fit temperature of the thermal component is $\sim4$keV for
the power-law model, while it is $\sim1.5$keV for the SRCut and cut-off
power-law models.  
The best-fit roll-off frequency of the SRCut model appears at around
0.9 keV, which is roughly consistent with that measured using the
Beppo-SAX spectrum (1.2 keV; \cite{Vink03}).  
The best-fit flux
($3.7\times10^{3}$ Jy) at 1 GHz for the SRCut model exceeds 
the measured radio flux by almost 40\% ($2.7\times10^{3}$ Jy: \cite{Green04}).
The overestimate of the model might be due to the non-linear acceleration effects (\cite{Atoyan00}).
We adopted the best-fit
parameters for the cut-off power-law model in the analysis below. 

The best-fit values shown in Table~\ref{bestfit} are dependent on the
effective area and relative normalization factor that we adopted. To
calculate the XIS effective area, we consider two cases for the angular
extension : a point-like source and the Chandra 4--6 keV band
image. For the two effective area, we further gave conservative
uncertainty of about 10\% (105--125\%) on the normalization constant.
In any case, the thermal-to-nonthermal fraction $f$ ranges from
0.1 to 0.2 for the cut-off power-law model. The $\Gamma$ and the
cut-off energy is sensitive to the combination of the effective area
and the normalization constant. If we make an assumption of
$\Gamma\sim2.3$, predicted by the classical diffusive shock
acceleration (DSA) theory (see \cite{Uchiyama07}), the best-fit cut-off energy always appears around 3 keV which is consistent with $>$ 1 keV.
The conclusion of this paper does not depend on these uncertainties. 
We thus adopted the best-fit value shown in Table~\ref{bestfit} in the discussion section. 


One may be interested in the flux difference between the HXD PIN and
the hard X-ray instruments onboard the other satellites. The HXD PIN
flux is $\sim$107\%, $\sim$126\%, and $\sim$133\% of those reported
using the data of Rossi-XTE (\cite{Allen97}), Beppo-SAX
(\cite{Favata97}), and Integral (\cite{Renaud06}), respectively.  To
calculate the ratio, we set the relative normalization factor to the
XIS as unity.  We assumed the best-fit model published in each paper
and fitted the HXD PIN data in the 15--40 keV band with the model but
multiplied it by the ratio.  The difference of the flux among the
instruments will be due to the systematic uncertainty with
unidentified reasons.


\begin{table}[ht]
  \caption{Best-fit parameters of the overall spectrum.} 
  \label{bestfit}
  \begin{center}
  \begin{tabular}{lcccccccc}
    \hline
 & \multicolumn{2}{c}{Thermal model} & \multicolumn{4}{c}{Non-thermal model} & $f^\ddagger$ & $\chi^2$/d.o.f.\\ \hline
 \multicolumn{7}{l}{ Thermal bremsstrahlung + Power-law } & \\  
 &    $kT$    & $F_{\rm thermal}$$^\dagger$            & $\Gamma$ & & Norm & $F_{\rm non-thermal}$$^\dagger$ &    \\         
 &  [ keV ]       & [ erg/s/cm$^2$ ] &   &    & [ $^{\ast}$ ]  & [ erg/s/cm$^2$ ]      &  \\ 
 & $4.0_{-0.7}^{+0.9}$ & $0.53\times10^{-10}$             & $3.06\pm0.05$ & & $1.09_{-0.12}^{+0.11}$ & $2.3\times10^{-10}$ & 0.2 & 861.7/751 \\ \hline
 \multicolumn{7}{l}{ Thermal bremsstrahlung + SRCut } & \\  
 &    $kT$    & $F_{\rm thermal}$$^\dagger$            & $\alpha$ & Roll-off & Norm & $F_{\rm non-thermal}$$^\dagger$ &    \\  
 &  [ keV ]       & [ erg/s/cm$^2$ ] &   & [ keV ]   & [ Jy at 1 GHz ] & [ erg/s/cm$^2$ ]      &  \\ 
 & $1.5_{-0.1}^{+0.2}$ & $0.28\times10^{-10}$             & $0.77$(fix) & $0.92^{+0.05}_{-0.08}$ & $3654_{-81}^{+76}$ & $2.6\times10^{-10}$ & 0.1 & 857.8/751 \\ \hline
 \multicolumn{7}{l}{ Thermal bremsstrahlung + Cut-off power-law } & \\  
 &    $kT$    & $F_{\rm thermal}$$^\dagger$            & $\Gamma$ & Cut-off & Norm & $F_{\rm non-thermal}$$^\dagger$ &    \\  
 &  [ keV ]       & [ erg/s/cm$^2$ ] &   & [ keV ]   & [ $^{\ast}$ ]  & [ erg/s/cm$^2$ ]      &  \\ 
 & $1.57_{-0.26}^{+0.43}$ & $0.28\times10^{-10}$             & $2.2\pm0.7$ & $3.4^{+\infty}_{-1.7}$ & $0.97_{-0.40}^{+0.33}$ & $2.6\times10^{-10}$ & 0.1 & 861.1/750 \\ \hline
    \hline
    \multicolumn{4}{@{}l@{}}{\hbox to 0pt{\parbox{75mm}{\footnotesize
	  \par\noindent

The errors are at 90\% confidence level. \\
Cut-off power-law model : $\epsilon^{-\Gamma} \exp\left[-\sqrt{\epsilon/\epsilon_c}\right]$ \\
Fit range: XIS: 3.4--3.6 and 4.2--6 and 8--14 keV, HXD: 15--40 keV.\\
\footnotemark[${\ast}$]  ph~keV$^{-1}$~cm$^{-2}$ at 1~keV \\
\footnotemark[$\dagger$] ergs$^{-1}$~s$^{-1}$~cm$^{-2}$ in the 4--10 keV \\
\footnotemark[$\ddagger$] $f=\frac{F_{\rm thermal}}{F_{\rm thermal}+F_{\rm non-thermal}}$
	}\hss}}     
  \end{tabular}
  \end{center}
\end{table}

\subsection{K-shell transition lines from Iron (Fe K$\alpha$ and K$\beta$)}

\begin{figure}[!htb]
\includegraphics[width=12cm,angle=270]{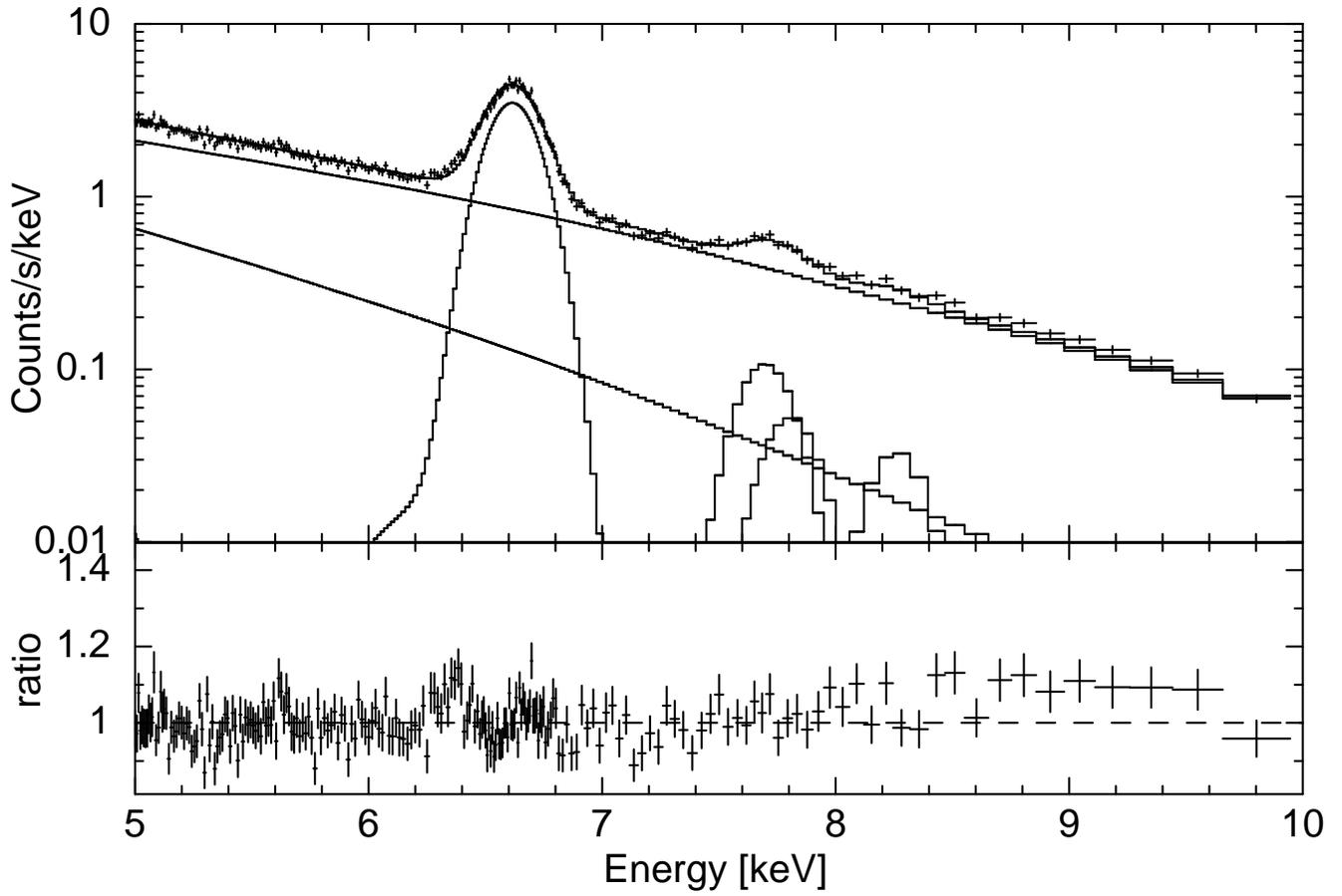}
\caption{Gaussian fit to the Fe/Ni K-lines. The best-fit parameters are listed in Table~\ref{4gauss}.
The thermal and non-thermal components of the model are also drawn with solid lines. The non-thermal component dominates the continuum flux.}
\label{FeK}
\end{figure}

The line-flux ratio $F_{\rm Fe-K\beta}/F_{\rm Fe-K\alpha}$ is
sensitive to the electron temperature. Figure~\ref{FeK} shows the zoomed-up XIS spectrum in the 5-10 keV
band. The strong line appears at 6.6 keV and is identified with
Fe-K$\alpha$ lines from highly ionized ions (helium-like or less
ionized). We also identified a blend of the Ni-K$\alpha$ and Fe-K$\beta$
lines at 7.6--7.9 keV and possibly the Fe-K$\gamma$ line at
$\sim8.3$~keV.  The blend appears as a broad line due to (1) the
Doppler shift and broadening, (2) the multiple lines from ions at
different ionization degrees (satellite lines), and (3) the limited
energy resolution of the XIS detector.

In order to determine the line fluxes of the blend, we fixed the line
energy of the Fe-K$\beta$ line to that of the Ni-K$\alpha$ multiplied
by a factor of 1.014 (e.g., \cite{koyama07b}). The best-fit parameters
are also shown in Table~\ref{4gauss}.  The upper and lower limits of the
flux ratio of $F_{\rm Fe-K\beta}/F_{\rm Fe-K\alpha}$ correspond to
the case that the line complex at 7.6--7.9 keV is dominated by
Fe-K$\beta$ and Ni-K$\alpha$, respectively. 

\begin{figure}[!htb]
\includegraphics[width=18cm,angle=270]{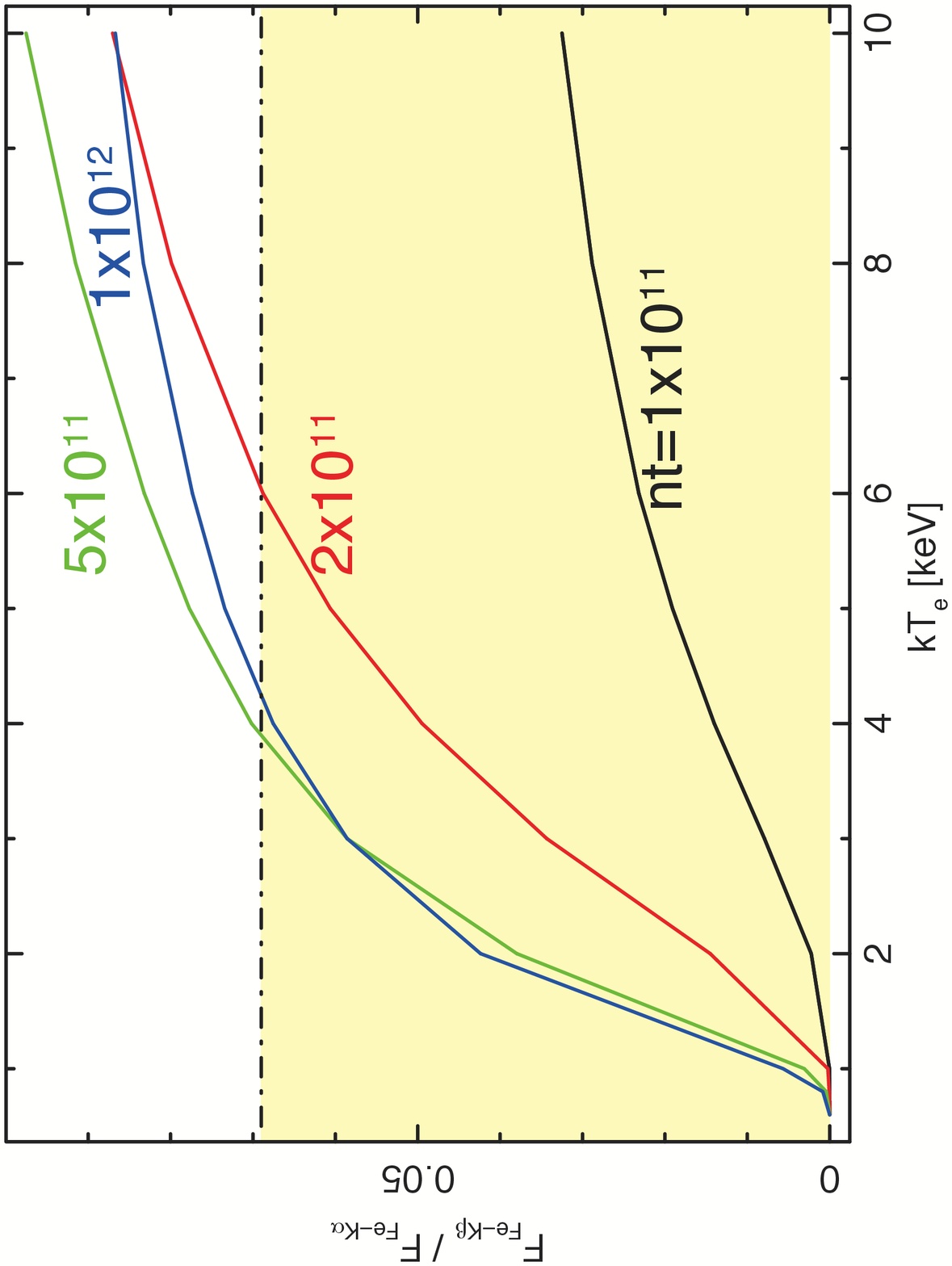}
\caption{Flux ratio of Fe-K$\alpha$ to Fe-K$\beta$ lines for a given electon tempereture (solid line). The yellow area is the best-fit ratio obtained from the line modelling (see text). We used the nei code in which the ionization equilibrium collisional plasma is assumed. The unit of the parameter $nt$ is s cm$^{-3}$. }
\label{lineratio}
\end{figure}

In Figure~\ref{FeK}, we compared the best-fit flux ratio $F_{\rm
  Fe-K\beta}/F_{\rm Fe-K\alpha}$ to those calculated at a given
temperature using the optically thin-thermal plasma code in
non-ionization equilibrium condition (the ``nei'' model in xspec fitting
package).  We faked the spectra for a pure-iron plasma with the
temperature of 0.8, 1, 2, 3, 4, 5, 6, 8, 10 keV and determined the
flux ratio by three-Gaussians with a thermal bremsstrahlung
continuum. We linearly interpolate the ratio, and make the plot shown in
Figure~\ref{lineratio}.

The best-fit K$\beta/$K$\alpha$ flux ratio can be explained if the
electron temperature is 1--4 keV in or near the collisional equilibrium
($nt=10^{12}$ s cm$^{-3}$). For the case of the short ionization
timescale of $1\times10^{11}$ s cm$^{-3}$, the wide range of electron
temperature over 10 keV is acceptable. The electron temperature
determined by the continuum modeling ranges from 1 to 5 keV for the
three different models. 
We conclude that the temperature derived from the Fe-line ratio is not inconsistent with the temperature derived above from the continuum models (1--5 keV depending on the model chosen).

\begin{table}[ht]
  \caption{Fe and Ni-lines from \casa.} 
 \label{4gauss}
  \begin{center}
  \begin{tabular}{lcc}
    \hline \hline
 &  [ keV ]       & [ $10^{-3}$ ph/s/cm$^2$] \\ \hline
Fe-K$\alpha$ &  $6.621\pm0.002$         & $4.47\pm0.07$ \\   
Ni-K$\alpha$ & $7.71_{-0.09}^{+0.04}$   & $<0.31$        \\
Fe-K$\beta$ & $7.81_{-0.09}^{+0.04}$$^{\ast}$ &  $<0.31$          \\
Fe-K$\gamma$ & 8.27(fix) &   $0.08_{-0.02}^{+0.05}$ \\ \hline
 \end{tabular}
  \end{center}
\footnotemark[${\ast}$] The line energy of Fe-K$\beta$ is linked to 1.014 times as high as that of Ni-K$\alpha$.\\
Fit range: XIS: 3.4--3.6 and 4.2--14 keV, HXD: 15--40 keV.\\
\end{table}

\subsection{K-shell transition line from Chromium (Cr K$\alpha$)}
\begin{figure}[!htb]
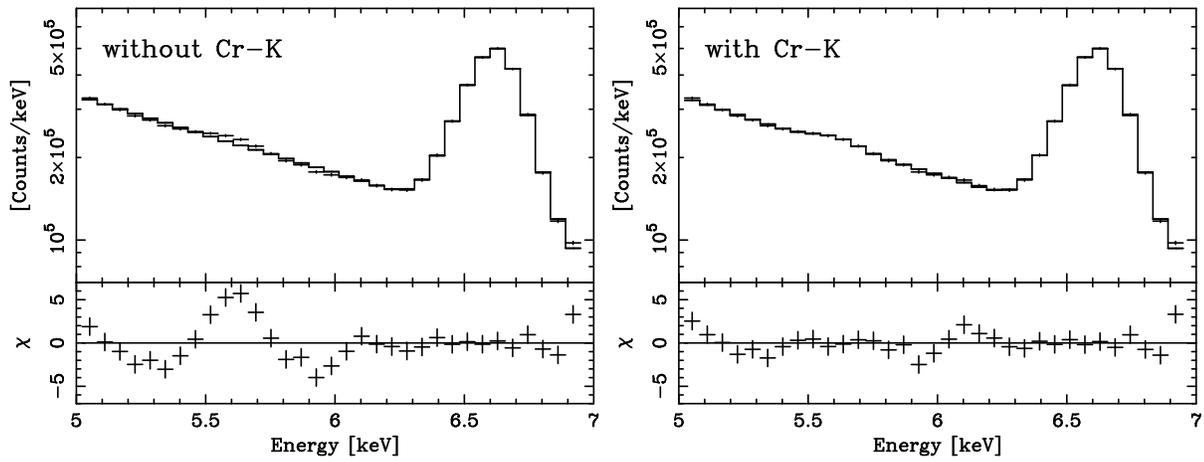

\includegraphics[width=6cm,angle=270]{figure8a.ps}
\includegraphics[width=6cm,angle=270]{figure8b.ps}
\caption{Spectrum from 5keV to 7keV.  Left: best fit results with a power law and Gaussian (Fe-K) and residual.  Right: best fit results with adding an extra Gaussian at 5.61keV and residual. }\label{CrK}
 \label{Cr}
\end{figure}

Figure~\ref{Cr} shows the zoomed-up XIS spectrum in the 5.0-7.0 keV
band. 
The residuals in the fitted spectrum shows a signature of Cr K-shell
emission. 
This weak signature was only found after the accumulation of
all available data of the 1st and 3rd observations, and by
adding up the data of four CCDs. Furthermore, in order to
bring out the signal in Figure~\ref{Cr} we used a courser
binning than in  Figure~\ref{xisspec}.

Since the energy band below 5\,keV may be affected by the group of
Ca-K emission line, we selected the energy band (5-7\,keV) to look for
a weak emission line.  
We fitted the data in the 5--7 keV band by using a power law and Gaussian line for Fe-K
profile.
The residual indicates a weak line-like feature around 5.6
keV.  We therefore introduce a Gaussian line around 5.6 keV that can
be a Cr-K$\alpha$ line.  In this way, we modeled the continuum with a
single cut-off power-law model and two lines with Gaussian functions:
one for the possible Cr-K$\alpha$ line and the other for the
Fe-K$\alpha$ line. Due to the limited statistics, we assumed the same
line broadening, i.e., Gaussian sigma, for both lines.  The best-fits and their residuals around the Cr-K$\alpha$ band are shown in Figure~\ref{CrK}.
We found the F-test probability of non-presence of Cr-K$\alpha$ is around 10$^{-6}$.
The best-fit Gaussian energies are $5.61\pm0.02$ keV for Cr-K$\alpha$
and $6.615\pm0.002$ keV for Fe-K$\alpha$, respectively. 
\footnote{
A systematic error due to the calibration uncertainties of the absolute energy
  gain is an order of 0.01 keV at 6 keV. 
In fact, the best-fit energy of the Fe-K$\alpha$ line differs by $\sim$6~eV between the two different dataset (Tables~\ref{4gauss} and \ref{Cr-line}). 
However, the discussion we made later is independent on the calibration error of the absolute energy scale. }

\section{Discussion}

\subsection{Non-thermal emission}

We have obtained a 3.4--40 keV continuum spectrum 
summed over the entire remnant with Suzaku.
It is now recognized that the continuum emission of \casa\ 
in this bandpass is comprised mainly of the two components: 
(1) thermal bremsstrahlung by shock-heated electrons 
mostly in the reverse-shocked ejecta, 
and 
(2)  synchrotron radiation produced by multi-TeV electrons.
Recent studies emphasize that 
the synchrotron component also should originate predominantly in the 
reverse-shocked ejecta (\cite{Uchiyama08a}; \cite{Helder08}), 
suggesting an interesting possibility that multi-TeV particles can be accelerated 
in supernova reverse shocks. 

Not only synchrotron radiation but also the non-thermal bremsstrahlung
was proposed to contribute to the non-thermal compornent
\citep{Laming01}. However, recent studies reported facts against the
non-thermal bremsstrahlung: the lack of line-emission in the continuum
areas \citep{Helder08}, Compton-cooling \citep{Vink08}, and fast
variability in the knots \citep{Uchiyama08a}. We think the contribution
of the non-thermal bremsstrahlung is minor.

We confirm the hard X-ray enhancement in 
the west ("the western spot") as shown in the 11--14 keV XIS image
(Fig.\ 2e).
The non-thermal continuum dominance in the west was previously reported by 
a 9--11 keV image made with Beppo-SAX (\cite{Helder08}) 
and a 8.1--15 keV image with XMM-Newton (\cite{Bleeker01}).
\citet{Helder08} argued that in the western part of \casa\, most 
X-ray synchrotron originates in the reverse shocks, and  
identified the reason for the synchrotron enhancement as 
a locally higher reverse shock velocity of $v_s \sim 6000\ \rm km\ s^{-1}$.

Interestingly, we found a hint that the peak of TeV $\gamma$-rays measured by
HEGRA and MAGIC coincides with the location of the synchrotron-dominated
western spot.  This suggests that the TeV $\gamma$-rays also can originate
from reverse-shocked ejecta.
The possible coincidence of bright 
X-ray continuum emission with TeV emission 
reinforces the idea that the X-ray continuum in the west is
indeed synchrotron radiation, 
irrespective of the TeV radiation mechanism. 
\citet{Atoyan00},\citet{Vink03} and \citet{Berezhko04} argued that the multi-TeV emission is likely to have an 
hadronic origin. Our finding of a possible keV-TeV correlation suggests
that the accelerated multi-TeV hadrons coul d be heavy elements (i.e.,
ejecta) accelerated by the reverse shock.

The spatially most
detailed information about the X-ray continuum emission in \casa\ 
came from Chandra observations. 
The superb angular resolution ($\sim 1\arcsec$) 
of Chandra made it possible to 
fully resolve the arcsecond-scale filamentary or knotty structures 
as well as more diffuse emission.
The extreme brightness of \casa\ and long exposures allocated for 
the ACIS observations (\cite{Hwang04}) allows for 
meaningful spectral fitting of arcsecond-scale pixels. 
Based on pixel-by-pixel photon indices determined for the 4.2--6 keV continuum band 
using a power-law fit in each $4.9\arcsec \times 4.9\arcsec$ pixel, 
\citet{Helder08} have derived the total contribution of 
the synchrotron component in the 4.2--6 keV band as $r \sim 54$\%,
 on the assumption that the continuum shapes characterized by 
 $\Gamma > \Gamma_{\rm th} = 3.1$ 
are of thermal origin. 
The synchrotron flux relative to thermal bremsstrahlung depends on 
the choice of  $\Gamma_{\rm th}$. 
Using Figure 10 of \citet{Helder08},
$\Gamma_{\rm th} = 2.8$ and $\Gamma_{\rm th} = 3.4$ translate in 
 $r \sim 33$\% and $r \sim 72$\%, respectively.
The difficulties in disentangling the two types of continuum emission 
stem from the tantalizing fact that they can be co-spatial, as well as 
from the projection effects. 
Moreover, the limited spectral coverage of Chandra ($< 8$ keV) 
does not allow to perform spectral decoupling. 

A complimentary approach, which the present work takes, 
is measuring a wide band 
X-ray spectrum extending to hard X-rays in order to break the degeneracy 
between the thermal bremsstrahlung and synchrotron components. 
A drawback of this approach is a lack of 
imaging capability above 15 keV  to resolve the spatial structures of 
hard X-ray emission. 
We are bound to the analysis of a wide-band  spectrum integrated over 
the entire face of the remnant  with a loss of
morphological information.
Still, we would be able to obtain meaningful results since 
the variation of electron temperatures determined by line modeling 
seems reasonably small, 
ranging from 1--2 keV (e.g., \cite{Hwang03}; \cite{Patnaude07}).
Though this issue should be examined using Chandra data in future work,
we here assume that possible temperature distributions do not have a major 
impact on our discussion. 

Our spectral modeling of the overall Suzaku XIS+PIN spectrum in 3.4--40 keV 
(excluding emission lines) 
with a bremsstrahlung+power law (BR+PL) model yielded the best-fit 
temperature of $kT_e = 4.0^{+0.9}_{-0.7}$ keV and photon index 
of $\Gamma = 3.06\pm 0.05$ 
with a thermal-to-nonthermal ratio of $f = 0.2$ 
(see Table 3).
Compared with previous results on the temperature 
obtained with Chandra, typically 1--2 keV, 
the BR+PL model gives a too-high temperature for bremsstrahlung.
From Figure 2 of \citet{Stage06}, 
in line-dominated thermal inner ring in the north 
and east, thermal bremsstrahlung has a temperature $< 2$ keV in most cases. 
On the other hand, a bremsstrahlung+cut-off power-law (BR+CPL) model resulted in 
the best-fit  cut-off energy of 
$\epsilon_c > 1$ keV, with $f\sim0.1$.
Irrespective of the actual spectral shape, 
we can conclude that the synchrotron component (integrated over the remnant) 
in the X-ray band should be described by a convex-shaped continuum, 
not by a simple power law.
This idea is 
independently supported by 
the photon index map given by \citet{Helder08}.
The derived cut-off energy (assuming $f\propto \epsilon^{-\Gamma} \exp
-\sqrt{\epsilon/\epsilon_c}$) of $\epsilon_c > 1$ keV is
consistent with that expected from shock acceleration theory (with
Bohm diffusion) given a typical reverse shock speed $v_s$ of
$6000\ \rm km\ s^{-1}$, as well as the X-ray variability of
synchrotron filaments ($\epsilon_c\ {\rm keV} = 2.2\ \eta^{-1}
(v_s/6000\ {\rm km}\ {\rm s}^{-1})$ ; see \cite{Uchiyama07}).

An interesting consequence of the spectral decomposition discussed
above is a fairly small value of $f \sim $0.1--0.2.  This implies that even
a 4--6 keV image made with Chandra should be dominated by the
synchrotron component. The electron temperature of 1--2 keV
corresponds to $\Gamma\sim$6.1--3.8 in the 4.2--6 keV band.  If we
assume the filaments with $\Gamma>$3.8, equivalent to $kT < 2$ keV,
are occupied by the thermal emission, Figure 10 of \citet{Helder08} also shows 
a thermal-to-nonthermal ratio of $f \sim 0.1$,
which is consistent with our value.

The small value of $f \sim $0.1--0.2 means the equivalent widths 
after subtracting a synchrotron spectrum become quite large. 
In particular, 
the equivalent width of Fe-K$\alpha$ should be as high as $\sim$7--14 keV, 
requiring iron-rich ejecta for Fe-K$\alpha$ line emission. 
In the western regions, outside the bright ejecta ring, 
extreme iron-rich ejecta  were found
(iron composition by mass amounting to 90\%), 
that likely originates from the $\alpha$-rich freezeout process,
which occurred in the innermost ejecta layer (\cite{Hwang03}).
The very large equivalent width (EW) of Fe-K$\alpha$ inferred for 
the integrated Suzaku XIS spectrum may be understood if 
the Fe line comes in large part from $\alpha$-rich freezeout ejecta.
Indeed, the presence of such $\alpha$-rich freezeout ejecta 
is required by the observed $^{44}\rm Ti$ emission 
(\cite{Vink01}; \cite{Renaud06}).


\subsection{Cr-K line}

\begin{table}
\caption{Cr line in SNRs}
\label{Cr-line}
\begin{tabular}{ccccccl}
\noalign{\smallskip} \hline \hline \noalign{\smallskip}
 &  \multicolumn{2}{c}{Cr} & & \multicolumn{1}{c}{Fe} \\ \cline{2-3}\cline{5-5}
Target & line energy & flux & & line energy & $R^{*}$ & reference (satellite) \\
  &  (eV)  & 10$^{-5}$photons\,cm$^{-2}$ s$^{-1}$ & & (eV)  & E$_{Cr}$/E$_{Fe}$ &\\
\noalign{\smallskip} \hline \noalign{\smallskip}
W49B  & 5685$^{+20}_{-27}$ & 30$^{+8}_{-11}$ & & 6658$^{+3}_{-2}$ & 0.854$\pm$0.004 & \cite{Hwang00b} (ASCA) \\ 
        & 5660$\pm$10 & 25$\pm$4 & & & & \cite{Miceli06} (XMM-Newton) \\ 
Tycho & 5480$\pm$20 & 2.45$^{+0.48}_{-0.42}$ & & 6445$\pm$1 & 0.850$\pm$0.004 & \cite{Tamagawa08} (Suzaku) \\ 
\casa\ & 5611$\pm$16 & 0.9$\pm$0.1  & & 6615$\pm$1 & 0.848$\pm$0.002 & This work (Suzaku) \\ 
\noalign{\smallskip} \hline \noalign{\smallskip}
\end{tabular}
$*$ $R$:the line energy ratio of Cr-K$\alpha$ to Fe-K$\alpha$.
\end{table}

There are two reported SNRs from which the Cr line has been detected; W49B
(\cite{Hwang00b,Miceli06}) and Tycho (\cite{Tamagawa08}).
\casa, reported here, is the third one.  The line center energies are
summarized in table~$\ref{Cr-line}$ for Cr and Fe emission lines as
well as the energy ratio, $R$, between them.  Emission lines in W49B
are ionized to He-like ions while those in Tycho are in low ionization
state. 

If Cr and Fe are in neutral/He/H-like ion, $R$ is
0.845/0.848/0.851. $R$ is quite insensitive to the ionization state.
The similar argument can be valid not only for $R$, but also for the Doppler shifting. If the
expanding speeds of Cr and Fe are the same, $R$ is not sensitive to
their Doppler velocity at all. 

According to the nucleosynthesis theory (\cite{Woosley94, Woosley95}),
$^{50}$Cr is produced through explosive oxygen and silicon burning as
itself, while the most abundant isotopes of chromium, $^{52}$Cr and
$^{53}$Cr are produced through $^{52,53}$Fe decay.  The most abundant
isotope of iron, $^{56}$Fe is produced through $^{56}$Ni decay.  They
are generated in a similar location deep inside the star. Therefore,
they must have been ejected with similar explosion speed and have
experienced as similar ionization time scale. This is consistent with
our results that the observed value of $R$ is identical to each SNR
within statistical uncertainties (table~$\ref{Cr-line}$).  

After submitting our paper for publication, we became aware of an
independent paper by \citet{Yang09} on the same subject on the Cr-K
line.  Their conclusions are very similar to those presented here: Cr
and Fe are colocated. The $R$ value of $0.848\pm0.002$ we measured with
the Suzaku data for \casa\ is in very good agreement with theirs
of $0.850\pm0.001$ using the Chandra data.

\section{Summary}

\begin{enumerate}

\item  
Suzaku observed the young supernova remnant \casa, and detected the
broadband continuum from 3.4 to 40 keV. The 5-$\sigma$ upper limit in
the 150--500 keV band is $1\times10^{-9}$ erg/s/cm$^2$ by assuming the
photon index of unity.

\item  
The line-flux ratio $F_{\rm Fe-K\beta}/F_{\rm Fe-K\alpha}$, is
consistent with the electron temperature measured with the continuum
modelling, but within the large error. 

\item
We modeled the X-ray continuum spectrum in the 3.4--40 keV band summed
over the entire remnant. For a model, we assumed the simplest
combination of the single thermal bremsstrahlung and the single
non-thermal cut-off power-law models.
We found that the X-ray continuum in the 3.4--40 keV band is dominated
by non-thermal emission. The thermal-to-nonthermal fraction of the
continuum flux in the 4--10 keV band is $\sim$0.1--0.2.  The non-thermal
continuum in the 3.4--40 keV band is reproduced with a cut-off
power-law model with a cut-off energy of $> 1$~keV. The
cut-off energy can be understood by the diffusive shock acceleration
at a shock velocity of as high as $\sim6000$ km s$^{-1}$.

\item  
We report on the detection of 
Cr-K$\alpha$ at 5.61 keV in \casa. \casa\ is the third object
for which the Cr-K$\alpha$ line emission has been detected. The elements Cr as well
as Fe are consistent with being generated in the  region deep inside the supernova, since they
seem to have had similar ionization history.

\item
A hard X-ray image of the continuum emission was obtained with the
XIS. Its brightest spot appears at the western part of the reverse
shock.  The XIS spectrum below 14 keV is smoothly connected to that of
the non-imaging detector HXD covering the energy band above 15
keV. Therefore, the 10 keV band image of the XIS is likely the good
approximation of the image above 10 keV.

\item
We also found a hint that the hard X-ray peak possibly coincides with
the TeV peak detected by HEGRA and MAGIC. Since the TeV emission is
likely to be hadronic origin, the possible keV-TeV correlation
suggests that the TeV hadron could be presumably accelerated in a
reverse shock. This is the first observational hint that the high-energy 
hadrons as well as leptons can be accelerated in the reverse
shock in a supernova remnant. A tighter constaraint of the TeV
position with deep exposures will be crutial to test this hint.

\end{enumerate}
\bigskip
We would like to express our sincere thanks to Prof. Gerd P{\"u}hlhofer for his insightful comments. 
We thank Dr. Javier Rico and his MAGIC Collaborators who kindly
provide the TeV data and technically guide us how to handle the
data. Prof. Katsuji Koyama gave us very useful comments on line analysis.
We also thank all members of the Suzaku team.  
EH and JV are supported by the Vidi grant of JV from the Netherlands Organization for Scientific Research (NWO).
This work is
partly supported by a Grant-in-Aid for Scientific Research by the
Ministry of Education, Culture, Sports, Science and Technology
(21018009 \& 16002004).



\begin{thebibliography}{}
\bibitem[Aharonian et al.(2001)]{Aharonian01} Aharonian, F., et al.\ 2001, \aap, 370, 112 


\bibitem[Aharonian et al.(2004)]{Aharonian04} Aharonian, F.~A., et
al.\ 2004, \nat, 432, 75

\bibitem[Aharonian et al.(2007)]{Aharonian07} Aharonian, F., et
al.\ 2007, \apj, 661, 236

\bibitem[Albert et al.(2007)]{Albert07} Albert, J., et al.\ 2007, \aap, 474, 937 

\bibitem[Allen et al.(1997)]{Allen97} Allen, G.~E., et al.\ 1997, \apjl, 487, L97 

\bibitem[Amenomiri et al. (2006)]{Tibet06} Amenomiri, M., et al. (Tibet AS{$\gamma$} Collaboration) 2006, Physics Letters B, 632, 58

\bibitem[Atoyan et al.(2000)]{Atoyan00} Atoyan, A.~M., Aharonian, F.~A., Tuffs, R.~J., V\"olk, H.~J.\ 2000, \aap, 355, 211 

\bibitem[Bamba et al.(2003)]{Bamba03} Bamba, A., Yamazaki, R., 
Ueno, M., \& Koyama, K.\ 2003, \apj, 589, 827 

\bibitem[Bamba et al.(2005)]{Bamba05} Bamba, A., Yamazaki, R., 
Yoshida, T., Terasawa, T., \& Koyama, K.\ 2005, \apj, 621, 793 

\bibitem[Berezhko {\& V\"{o}}lk(2004)]{Berezhko04} Berezhko, E.~G., {\ V\"{o}}lk, H.~J.\ 2004, \aap, 419, L27 


\bibitem[Bleeker et al.(2001)]{Bleeker01} Bleeker, J.~A.~M., Willingale, R., van der Heyden, K., Dennerl, K., Kaastra, J.~S., Aschenbach, B., \& Vink, J.\ 2001, \aap, 365, L225 


\bibitem[Favata et 
al.(1997)]{Favata97} Favata, F., et al.\ 1997, \aap, 324, L49 

\bibitem[Fukazawa et al.(2009)]{Fukazawa09} Fukazawa, Y., et al. 2009, \pasj, 61, 17 


\bibitem[Green(2004)]{Green04}
Green, D.~A.\ 2004,
A Catalogue of Galactic Supernova Remnants (2004 January version), 
(Cambridge, UK, Mullard Radio Astronomy Observatory)
available on the WWW at http://www.mrao.cam.ac.uk/surveys/snrs/

\bibitem[Helder 
\& Vink(2008)]{Helder08} Helder, E.~A., \& Vink, J.\ 2008, ApJ, 686, 1094


\bibitem[Holt et al.(1994)]{Holt94} Holt, S.~S., Gotthelf, 
E.~V., Tsunemi, H., \& Negoro, H.\ 1994, \pasj, 46, L151 


\bibitem[Hughes et al.(2000)]{Hughes00} Hughes, J.~P., Rakowski, 
C.~E., Burrows, D.~N., \& Slane, P.~O.\ 2000, \apjl, 528, L109 


\bibitem[Hwang et al.(2000)]{Hwang00b}
Hwang,~U., Petre,~R., Hughes,~J.~P.\ 2000, \apj, 532, 970

\bibitem[Hwang 
\& Laming(2003)]{Hwang03} Hwang, U., \& Laming, J.~M.\ 2003, \apj, 597, 362 

\bibitem[Hwang et al.(2004)]{Hwang04} Hwang, U., et al.\ 2004, 
\apjl, 615, L117 

\bibitem[Ishisaki~\etal(2007)]{ishisaki07} Ishisaki, Y.,~\etal\  2007, \pasj, 59, S113

\bibitem[Kokubun~\etal(2007)]{kokubun07} Kokubun, M.,~\etal\
  2007, \pasj, 59, S53

\bibitem[Koyama et al.(1995)]{Koyama95} Koyama,~K., et al. 1995, \nat, 378, 255

\bibitem[Koyama~\etal(2007a)]{koyama07a} Koyama, K.,~\etal \
  2007, \pasj, 59, S23

\bibitem[Koyama~\etal(2007b)]{koyama07b} Koyama, K.,~\etal \
  2007, \pasj, 59, S245

\bibitem[Laming(2001)]{Laming01} Laming, J.~M.\ 2001, \apj, 546, 1149 

\bibitem[Laming \& Hwang(2003)]{Laming03} Laming, J.~M., \& Hwang, U.\ 2003, \apj, 597, 347 

\bibitem[Long et al.(2003)]{long03} Long, K.~S., Reynolds, S.~P., Raymond, J.~C., Winkler, P.~F., Dyer, K.~K., \& Petre, R.\   2003, \apj, 586, 1162

\bibitem[Miceli et al.(2006)]{Miceli06}
Miceli,~M, \etal\ 2006, \aap, 453, 567

\bibitem[Mitsuda~\etal(2007)]{mitsuda07} Mitsuda, K.,~\etal\
  2007, \pasj, 59, S1

\bibitem[Muraishi et al.(2000)]{muraishi00} Muraishi, et al.\
(CANGAROO collaboration) 2000, \aap, 354, L57

\bibitem[Parizot et 
al.(2006)]{Parizot06} Parizot, E., Marcowith, A., Ballet, J., \& Gallant, Y.~A.\ 2006, \aap, 453, 387 

\bibitem[Patnaude 
\& Fesen(2007)]{Patnaude07} Patnaude, D.~J., \& Fesen, R.~A.\ 2007, \aj, 133, 147 

\bibitem[Pravdo 
\& Smith(1979)]{Pravdo79} Pravdo, S.~H., \& Smith, B.~W.\ 1979, \apjl, 234, L195 

\bibitem[Renaud et al.(2006)]{Renaud06} Renaud, M., et al.\ 
2006, \apjl, 647, L41 

\bibitem[Serlemitsos et al.(1973)]{Serlemitsos73} Serlemitsos, P.~J., 
Boldt, E.~A., Holt, S.~S., Ramaty, R., 
\& Brisken, A.~F.\ 1973, \apjl, 184, L1 

\bibitem[Serlemitsos~\etal(2007)]{serlemitsos07} Serlemitsos, P.,~\etal
  \ 2007, \pasj, 59, S9

\bibitem[Stage et al.(2006)]{Stage06} Stage, M.~D., Allen, 
G.~E., Houck, J.~C., \& Davis, J.~E.\ 2006, Nature Physics, 2, 614 

\bibitem[Takahashi~\etal(2007)]{takahashi07} Takahashi, T., et
  al.\ 2007, \pasj, 59, S35

\bibitem[Tamagawa et al.(2008)]{Tamagawa08} Tamagawa, T., et al.\ 
2008, \pasj, accepted (arXiv:0805.3377)

\bibitem[Uchiyama et al.(2007)]{Uchiyama07} Uchiyama, Y., 
Aharonian, F.~A., Tanaka, T., Takahashi, T., 
\& Maeda, Y.\ 2007, \nat, 449, 576 

\bibitem[Uchiyama 
\& Aharonian(2008)]{Uchiyama08a} Uchiyama, Y., \& Aharonian, F.~A.\ 2008, \apjl, 677, L105 

\bibitem[Uchiyama et al.(2008)]{Uchiyama08b} Uchiyama, Y., et al.\ 2008, \pasj, 60, 35 

\bibitem[Vink et al.(2001)]{Vink01} Vink, J., Laming, J.~M., 
Kaastra, J.~S., Bleeker, J.~A.~M., Bloemen, H., 
\& Oberlack, U.\ 2001, \apjl, 560, L79 

\bibitem[Vink 
\& Laming(2003)]{Vink03} Vink, J., \& Laming, J.~M.\ 2003, \apj, 584, 758 

\bibitem[Vink(2008)]{Vink08} Vink, J.\ 2008, \aap, 486, 837 

\bibitem[Willingale et al.(2002)]{Willingale02} Willingale, R., Bleeker, J.~A.~M., van der Heyden, K.~J., Kaastra, J.~S., \& Vink, J.\ 2002, \aap, 381, 1039 

\bibitem[Woosley et al.(1994)]{Woosley94} Woosley, S.~E., Eastman, R.~G., Weaver, T.~A., \& Pinto, P.~A.\ 1994, \apj, 429, 300 

\bibitem[Woosley \& Weaver(1995)]{Woosley95} Woosley, S.~E., \& Weaver, T.~A.\ 1995, \apjs, 101, 181 

\bibitem[Yang et al.(2009)]{Yang09} Yang, X.~J., Tsunemi, H., 
Lu, F.~J., \& Chen, L.\ 2009, \apj, 692, 894 


\end{thebibliography}
\end{document}